\documentclass[a4paper,english,pra,twocolumn]{revtex4}

\usepackage{amsfonts}
\usepackage{babel}
\usepackage{graphicx}
\usepackage{amssymb}
\usepackage{revsymb}

\begin{document}
\title{Quantumlike Diffusion over Discrete Sets}
\author{Demian Battaglia}
\affiliation{Dipartimento di Fisica Teorica, Universit\`a di Torino, 10130 Torino, 
Italy,  \\ and\\  SISSA, I-34014 Trieste, Italy}
\author{Mario Rasetti}
\affiliation{Dipartimento di Fisica and Unit\`a INFM, \\ Politecnico di Torino, I-10129 
Torino, Italy}
\date{\today}
\begin{abstract}

\hfil{\bf Abstract}\hfil
 \\ \par
In the present paper, a discrete differential calculus is introduced and 
used to describe dynamical systems over arbitrary graphs. The discretization of space 
and time allows the derivation of Heisenberg-like uncertainty inequalities and of a 
Schr\"{o}dinger-like equation of motion, without need of any quantization procedure.
\end{abstract}
\maketitle

Ever since the seminal work of Regge \cite{RE} on gravity, discretization has 
been often used in classical physics to explore complex continuous theories in a 
noperturbative fashion. The idea there is to reduce the infinite number of 
degrees of freedom of (typically Riemannian) manifolds by dealing instead with 
piecelinear spaces described by a finite number of parameters. For example, 
in standard Regge calculus a manifold is approximated by a simplicial lattice 
with fixed coordination number, while in the dynamical triangulated random surface 
method \cite{KA} the very coordination numbers are treated as dynamical degrees 
of freedom. The notion of discrete ambient space (or space-time) has thus become one 
of the standard ways of making complex geometries and topologies accessible to 
(essentially combinatorial) discrete calculations, in a setting naturally 
renormalized.     

On the other hand, when quantum features are considered, discreteness plays 
an unexpected role. It was shown that merely importing conventional quantum 
models on ambient spaces that are represented by inhomogeneous graphs \cite{BEC-1}, 
\cite{BEC-2} the graph inhomogeneity plays the role of curvature, giving rise to 
effective interactions among free particles. The question raises of course of whether 
or not a quantum model of this sort of a system living on a graph, represented 
e.g. in the second quantization formalism, actually describes the real 
(possibly interacting) physical system we intend to represent; whether or not 
is has a classical counterpart from which it can be obtained by a process of 
quantization. Canonical quantization, however, requires an underlying structure 
that is not manifestly evident when the system lives on a graph.        

In this paper we intend to explore some of the structural features that 
underly the problem. 
 
Let us consider a dynamical system whose ambient space, discrete, is provided by a
graph $X$, either finite or infinite. The graph will be assumed to be completely defined 
by its vertex and edge sets, $V(X)$ and $E(X)$, respectively. The edge connecting two 
given vertices $v$ and $v'$ will be denoted as $\langle vv'\rangle$. The geometry of
$X$ arises completely from the adjacency relations encoded in the adjacency matrix 
$\mathbb{A}(X)$, whose entry $A_{vv'}\, ; \, v,v' \in V(X)$ is different from 0 and 
equal to 1 if and only if $v$ is adjacent to $v'$, {\sl i.e.} if $\langle vv'\rangle \in 
E(X)$. It should be noticed that in ordinary differential geometry, graphs, like any 
other point set, can be considered as 0-dimensional real (or complex) manifolds. The 
adjacency matrix $\mathbb{A}(X)$ is related to the topology of $X$: a classical example 
is provided by comparison between complete graphs $\mathsf{K}_6$, which can be embedded 
planarly in the projective plane, but not in the euclidean one, and $\mathsf{K}_7$, 
which can be embedded in the 2-torus. In this sense it is possible to say that the graph geometry has a 
combinatorial origin. Nevertheless, a graph $X$ is still a topological space, 
even if not embedded in some differential manifold: it can indeed be endowed with the 
topology induced {\sl e.g.} by the chemical distance metric (the chemical distance between
$v$ and $v'$ being the number of edges of the shortest path connecting the two vertices). 
Moreover an atlas of $X$ over $\mathbb{Z}$ can be defined, simply by fixing an integer
labeling of the vertices of the graph. In that case it would be possible to consider $X$ 
as a sort of one dimensional \emph{integer manifold}, with a non trivial topology related 
to its connectivity properties and with every vertex unambiguously identified by one 
integer coordinate.

Let us now introduce the commutative algebra $\mathcal{A}(X)$ of complex valued functions 
over $X$, together with the ordinary function product. A basis is provided by the set of 
vertex characteristic functions $\{a_{v}| v\in V(X)\}$, such that $a_{v}(v')=\delta_{v,v'}$. If 
$f(v)=f_v$, then one can write $f=\sum_{v\in V(X)}f_{v}\;a_{v} \;$. 
$'$Orthogonality$'$ and $'$completeness$'$ relations hold: $a_{v}\cdot a_{v'}= \delta_{v,v'}\;a_{v} \; , \,\,\,
\sum_{v\in V(X)} a_{v} = \openone \;$, 
where $\openone$ is the multiplicative identity.

Another set of generators $\{a_{\langle vv'\rangle}|\langle vv'\rangle\in E(X)\}$ may be 
also defined, in one to one correspondence with unoriented edges $\langle vv'\rangle$. A 
generic $'$\emph{1-form}$'$ $\omega$, belonging to the set $\Omega(X)$, will be written as 
$\displaystyle{\omega = \sum_{\langle vv'\rangle \in E(X)} \omega_{\langle v'v\rangle} 
a_{\langle vv'\rangle}}$. Identification of $\omega$ as a 1-form is based on the 
existence of the following structure. First three different distributive product laws 
can be introduced:
\begin{equation}
\begin{array}{c}
a_{\langle vv'\rangle}a_{\langle ww'\rangle} = a_{\langle vv'\rangle}\; \delta_{v,w}\;
\delta_{v',w'} \; , \\
f\;a_{\langle vv'\rangle}=f_{v}\;a_{\langle vv'\rangle}\; , \; a_{\langle vv'\rangle}f 
= f_{v'}\;a_{\langle vv'\rangle} \; , 
\end{array}
\end{equation}
together with the operator $d$, connecting functions in $\mathcal{A}(X)$ to their 
differentials, 
\begin{equation}
\label{d1equaltozero}
a_{\langle vv'\rangle} = \left\{ \begin{array}{ll} a_{v}\; da_{v'}&v\neq v'\; 
\\ 0&v=v' \;  \end{array}\right. ;\,\,\,\,\,\,\, d(\openone)=0 \; .
\end{equation}
It follows from  (\ref{d1equaltozero}) that the 
generators $\{a_{\langle vv'\rangle}\}$ are not linearly independent, and, therefore, 
that different explicit expansions of fundamental differentials $\{da_{v} \in \Omega 
(X)\}$ can exist. It is evident  that $\displaystyle{df = \sum_{\langle vv'\rangle \in E(X)} f_{v'}\; 
a_{\langle vv'\rangle}}$, whence the following antisymmetrized expansion derives:
\begin{equation}\label{APG}
df = \sum_{\langle vv'\rangle \in E(X)}(f_{v'}-f_{v})a_{\langle vv'\rangle} \; .
\end{equation}
The two expressions differ only by a 1-form proportional to $d(\openone)$, null by 
definition. One should notice, however, that even if $d(\openone)$ is the identity for 
the sum of 1-forms, it is not a trivial multiplicative absorbing element, because the 
product of two null 1-forms is in general a 1-form different from 0. Let consider for 
instance $\displaystyle{\omega_{\openone} = \sum_{\langle vv'\rangle \in E(X)} a_{\langle
vv'\rangle} = \openone\; d(\openone) = 0}$: nevertheless one has $\omega_{\openone} f = \omega_{\openone} f - 0= \omega_{\openone} 
f - f \;\omega_{\openone}$, that is:
\begin{equation}\label{diffop}
\omega_{\openone} f = df \; , 
\end{equation}
which is not a null form unless $f$ is constant over $X$. Thus, the infinitely many 
possible expansions of the differential are not strictly equivalent one to the other and we 
have to fix a particular way of writing it. The choice (\ref{APG}) is the most natural 
because of its manifest similarity with a finite differences approximation of the usual 
differential. It should however be recalled that (\ref{APG}) is an exact relation: the 
ambient space in fact is discrete and there is no need to take continuous limits (in 
other words, we are dealing with a generalization of the $h$-calculus, see {\sl e.g.} 
\cite{kacquantumcalculus}).
Quite interestingly, the definition (\ref{APG}) yelds an unconventional form of the 
Leibnitz rule:
\begin{equation}\label{LeibnitzAPG}
d(fg) = fdg + gdf + df\, dg \; . 
\end{equation}
It can be easily checked that every expansion of the differential $df$ produces some 
additional nonlinear term in the Leibnitz rule: this is a characteristic feature of the 
differential defined by  (\ref{d1equaltozero}), and 
it is not present in other discrete calculi known in the literature (like, for instance, 
\cite{dimakis}).


A notion of discrete derivative can be introduced as well. A tangent \emph{vector field} 
\textbf{V} over $X$, belonging to the set $\chi(X)$, is a linear combination 
$\displaystyle{\mathbf{V}=\sum_{\langle vv'\rangle \in E(X)} V_{\langle vv'\rangle} 
\partial_{\langle vv'\rangle}}$ of the generators defined by means of the following 
$'$duality$'$ relation $\langle \cdot | \cdot\rangle : \Omega(X)\times \chi(X)\rightarrow 
\mathcal{A}(X)\;\,,\,
\langle a_{\langle vv'\rangle} | \partial_{\langle ww'\rangle}\rangle = \delta_{v,w} 
\delta_{v',w'} a_{v} \; $.
The action of a vector field on a scalar function is then:
\begin{equation}\label{vectorderivative}
\textbf{V}(f)= \langle df | \textbf{V}\rangle = \sum_{\langle vv'\rangle \in E(X)}
\left (f_{v'}-f_{v} \right ) V_{\langle vv'\rangle} a_{v} \; , 
\end{equation}
so that $\partial_{\langle vv'\rangle}$ can be interpreted as acting as a derivative along the directed edge $\langle vv'\rangle$: $\partial_{\langle vv'\rangle}(f)=(f_{v'}-f_{v}) a_{v} \;$.
In a sense the derivative along a vector field \textbf{V} can be considered as a derivative along a 
superposition of different edges $\langle vv'\rangle \in E(X)$ with weights given by $V_{\langle vv'\rangle}$.

Vector fields can also be interpreted as generators of transformations of $\mathcal{A}
(X)$, turning the set $\chi(X)$ into an algebra. Let $\mathfrak{m}$ be a map $\mathfrak{m} : \mathcal{A}(X) \rightarrow \mathcal{A} 
(X)$. Indeed setting  
\begin{equation}\label{infinitesimalgen}
\mathfrak{m} = \openone + \mathcal{V}_{\mathfrak{m}}
\end{equation}
one defines first $\mathcal{V}_{\mathfrak{m}}$, referred to as the \emph{infinitesimal 
generator} of $\mathfrak{m}$, which can be assumed, with no loss of generality, to belong 
to the algebra  of discrete derivatives along vector fields. The name is obviously 
suggested by Lie algebras theory, because of the analogy between (\ref{infinitesimalgen}) 
and the first-order perturbative expansion of an exponential. (\ref{infinitesimalgen}) is 
nevertheless once more an exact relation.
Consider now the endomorphisms associated to every single permutation 
$\pi$ of the vertices in $V(X)$: $\forall v \in V(X)\; ,\; \mathfrak{m}_{\pi}(a_{v}) = 
a_{\pi (v)}$; their group is a representation of the symmetric group $S_{p}$ acting over 
the algebra $\mathcal{A}(X)$, where $p=|X|$ is the order of the graph. Every $\mathfrak{m}_{\pi}$ 
has a vector field as infinitesimal generator: ${\bf V}_{\pi} = \sum_{v\in V(X)}\partial_{\pi(v), v} \;$ .
The vector fields $\textbf{V}_{\pi}$ are then the infinitesimal generators of the action 
of $S_p$ over $\mathcal{A}(X)$. Cayley's theorem, on the other hand, states that every 
finite discrete group is isomorphic to a suitable symmetric group; to show that, one might in general resort to the right adjoint action. Set:
\begin{equation}
\forall g \in \mathfrak{G}\; ,\; {\rm let}\; \mathcal{R}_{g}: \mathfrak{G} \rightarrow 
\mathfrak{G}\, ,\; \mathcal{R}_{g}(g')=g'g \; . 
\end{equation}
The $\{\mathcal{R}_{g}|g\in\mathfrak{G}\}$ form a representation of $\mathfrak{G}$ over $S_{|\mathfrak{G}|}$;
the infinitesimal generator of $g$ being then represented by the vector field:
\begin{equation}
{\bf V}_{g} = \sum_{v\in V(X)} \partial_{{\cal R}_{g}(v), v} \; . 
\end{equation}


Let now focus the attention on the union of the algebra $\mathcal{A}(X)$ with the space
of symmetric 1-forms: $T^{*}_{s}X=\mathcal{A}(X)\cup\Omega_{s}(X)$; and with the space of symmetric vector fields: $T_{s}X=\mathcal{A}(X)\cup\chi_{s}(X)$; a 1-form or a vector field are symmetric if  their components respect the relations $A_{\langle vv'\rangle}\equiv A_{\langle v'v\rangle}$. Both $T_{s}X$ and $T^{*}_{s}X$  can then be endowed with a symplectic structure by introducing suitable Poisson 
brakets; for instance, in the case of $T^{*}_{s}X$ one defines the bilinear operator $\{\cdot,\cdot\}: T^{*}_{s}X\times T^{*}_{s}X\rightarrow \mathcal{A}(X)$ such that:
\begin{equation}\label{poisson}
\begin{array}{l}
\{a_{v},a_{v'}\}=0  \\
\{a_{(\langle vv'\rangle)},a_{w}\}=(\delta_{v'w}-\delta_{wv})a_{v}\\
\{a_{(\langle vv'\rangle)}, a_{(\langle ww'\rangle)}\}=0
\end{array}
\end{equation}
The symmetrization operator is $a_{(\langle vv'\rangle)})=\frac{1}{2}(a_{\langle vv'\rangle}+a_{\langle v'v\rangle})$. A similar bracket can be introduced in $T_{s}X$  by substituting everywhere $a_{\langle vv'\rangle}$ with $\partial_{\langle vv'\rangle}$, obtaining then:
\begin{equation}\label{dynamicpoisson}
\{X,f\}=X(f)=\left<df|X\right>
\end{equation}
The operators defined in (\ref{poisson}) (and (\ref{dynamicpoisson}) ) are actually  Lie products, because they are antisymmetric and respect  the Leibniz and the Jacobi relations:
\begin{equation}\begin{array}{c}
\{f,\omega\}=-\{\omega,f\}\\
\{f g, \omega\}=\{f, \omega g\}+\{g,f\omega\}\\
\{\omega,\{\eta,f\}\}+\{\eta,\{f,\omega\}\}=0
\end{array}
\end{equation}
where $f, g\in\mathcal{A}(X)\,,\,\,\,X\in\chi_{s}(X)\,,\,\,\,\omega,\eta \in\Omega_{s}(X)$ (symmetricity is a sufficient condition for proving the Jacobi property).  It can then be 
stated that generators of functions and of 1-forms (vector fields) over the graph $X$ are canonically 
mutually conjugate; in a sense, one may claim that a canonical conjugation relation holds 
between degrees of freedom related to vertices and to edges of the graph. 

Consider now a material particle constrained to move over the graph $X$. At each time 
the particle will lie in one vertex $v \in V(X)$ and at the following step it will reach 
one of the sites $v'$ connected with $v$, such that $\langle vv'\rangle\in E(X)$ as 
specified by the adjacency matrix $\mathbb{A}(X)$. One can define a $'$state function$'$, say 
$\psi$, belonging to the function space $\mathcal{A}(X)$, whose meaning is different according to the dynamical description of the system one aims to achieve: if the particle position can be known 
precisely at each time step, its configuration (and hence its state) will be simply described by a vertex characteristic function $a_v$, which is non zero only at the site where the particle is 
located; if the particle is a classical random walker (or is a quantum particle) the state function (or its square modulus) should provide the 
corresponding localization probability density over $X$. On the other hand, the most general way to introduce a position operator is by defining it through a parameterization $\mathcal{Q}: V(X)\rightarrow \mathbb{Z}
\; ,\; v\mapsto q_{v}$ and an associated function $Q=\sum_{v\in V(X)} q_{v} 
a_{v}$. The action of the position operator on the state function will then simply be required to be:
\begin{equation}
\mathcal{Q}(\psi)=Q\cdot \psi \; . 
\end{equation}
In particular the vertex characteristic functions will be eigenstates of the position 
operator: $\mathcal{Q}(a_{v})=q_{v}a_{v}$. If the vertices of graph $X$ correspond to 
positions eigenstates, its edges will have to be connected to directions of motion, 
that is to velocities or momenta: this is the most natural physical interpretation of 
the canonical conjugation between functions and 1-form generators. If the particle hops 
from its initial position to a neighbouring site, it will cover always a single unit 
of chemical distance and every bond will be associated to the same elementary momentum 
eigenvalue, that can be set equal to 1 without loss of generality. An obvious choice 
for the momentum operator $\mathcal{P}$ will then be the symmetric 1-form $\omega_{\openone}$, which is such 
that $\omega_{\openone} a_{\langle vv'\rangle}= a_{\langle vv'\rangle}$:
\begin{equation}
\mathcal{P}(\psi )=\omega_{\openone} \psi = d\psi \; . 
\end{equation}
The Poisson parenthesis of position and momentum operator is then different from zero, and 
gives rise to canonical conjugation:
\begin{equation}\label{commutatore}
\{\mathcal{P},\mathcal{Q}\} = \Delta Q=\sum_{\langle vv'\rangle}(q_{v'}-q_{v})a_{v} \; . 
\end{equation}

The physical meaning of the momentum operator can be made more trasparent if the graph $X$ 
is endowed with a richer structure. Let $X$ be the Cayley graph of some group $\mathfrak G$ 
generated by a set $\mathfrak H$ of elements:  for example, if   $\mathfrak G$  is defined by a presentation, the set of the group generators provides a natural choice for $\mathfrak H$. A general function in $\mathcal{A}(X)$ will be 
written $\displaystyle{\psi =\sum_{g\in\mathfrak{G}} \psi_{g}\, a_{g}}$ and its differential 
as $\displaystyle{d\psi =\sum_{g\in\mathfrak{G}}\sum_{h\in\mathfrak{H}}(\psi_{gh}-\psi_{g})\,  
a_{\langle g~gh\rangle}}$. Let now $\widehat{X}$ be a graph whose vertices are in one-to-one 
correspondence with the complex numbers $\{\chi(h)\}$, where $\chi$ denote now the characters of 
the group $\mathfrak{G}$ (corresponding to its irreducible representations) and $h\in 
\mathfrak H$. This is consistent because the  set $\mathfrak{H}$ of generators represents the set of possible $'$directions of motion$'$ from each point of $X$: the vertices of the dual graph $\widehat{X}$, being labeled by the characters evaluated along the elements of $\mathfrak{H}$, are therefore associated to possible moves over $X$. One should of course choose different generators in different conjugacy 
classes, in order to have inequivalent $\chi(h)$.  On the contrary, there are no constraints 
on $E(\widehat{X})$ (whose element $\langle gg' \rangle\; ,\; g,g' \in {\mathbb{G}}$ 
depends on the single group element $g'' = g'\, g^{-1}$): nevertheless, if $\mathbb G$ is 
abelian, the set of its characters will be again a group and appropriate choices of the edge 
set will make $\widehat{X}$ one of its Cayley graphs. The group structure allows harmonic analysis over the algebra $\mathcal{A}(X)$; a $'$Fourier transform$'$ $\displaystyle{\widehat{\psi} = 
\sum_{\chi(h)} \psi_{\chi}\, b_{\chi(h)}}$ of $\psi$ can indeed be properly defined (both 
in the abelian and in the non abelian cases):
\begin{equation}\label{fourier}
\psi_{\chi} = \sum_{g\in {\mathfrak{G}}} \psi_{g} \chi^{*}(g) \quad ,\quad \psi_{g} = 
\frac{1}{GH} \sum_{\chi} \psi_{\chi}\, \chi(g) \; , 
\end{equation}
where $G$ and $H$ are the orders of $\mathfrak G$ and $\mathfrak H$, respectively (a different 
normalization constant will appear in the infinite group case). Like in the case of the
ordinary Fourier transform, Plancherel and Parseval theorems hold; given a function 
$t\in\mathcal{A}(X)$ such that $t_{g} = \psi_{gt}$, the components of its  Fourier 
transform will be $t_{\chi} = \chi(t)\, \psi_{\chi}$; and for any norm defined in
$\mathcal{A}(X)$ it will be possible to introduce one in $\mathcal{A}(\widehat{X})$
such that $\displaystyle{\|\psi \|^{2}=\frac{1}{GH}\|\widehat{\psi }\|^{2}}$.

Momentum can now be introduced as a sort of not injective parameterization over 
$\widehat{X}$, namely a linear operator defined in such a way that:
\begin{equation}
\widehat{\mathcal{P}}(b_{\chi(h)}) = p_{\chi(h)}\, b_{\chi(h)} \quad ,
\end{equation}
with $\chi(h) = 1 + p_{\chi(h)} \; $.
Similarity with the customary relation $e^{ikx}\sim 1 + ikx + {\cal O}(k^{2})$ should 
be noticed once more. 
Let now consider the case of a quantum particle. Expectation values of 
the square of position and of momentum are interesting because they give information about their variances: explicit calculation shows that uncertainty inequalities hold in full analogy with Heisenberg
Principle. In a state $\psi=\sum_{g\in 
{\mathfrak{G}}} \psi_{g}\, a_{g}$, the second probability distribution momenta are given by 
$\displaystyle{\langle\mathcal{Q}^{2}\rangle_{\psi}= \sum_{g\in{\mathfrak{G}}} |\psi_{g}|^{2}
q_{g}}$ and by $\displaystyle{\langle{\hat{\mathcal{P}}^2\rangle_{\psi} = \sum_{\chi,h\in 
{\mathfrak{H}}} \left |\psi_{\chi}\right |^2\, (\chi(h)-1)}}$. One has therefore: $
 \langle\mathcal{Q}^{2}\rangle_{\psi} \langle{\hat{\mathcal{P}}}^2\rangle_{\psi}
= \sum_{g\in{\mathfrak{G}}} \left |\psi_{g}\right |^{2}\, q_{g}^{2} \sum_{\chi, h\in 
{\mathfrak{H}}} \left |\psi_{\chi}\right |^2\, |\chi (h)-1|^2  \geq \frac{1}{H} \sum_{g \in{\mathfrak{G}}, h\in{\mathfrak{H}}} \left |\psi_{gh}\right |^2\, 
q_{gh}^2 \sum_{\chi,k\in{\mathfrak{H}}} \left |\psi_{\chi}\right |^2\, \left |\chi(k) - 1
\right |^2 \; $.
Once again, by Plancherel theorem $\left |\mathcal{P}(\psi )\right |^2 =\sum_{g\in{\mathfrak{G}},h\in{\mathfrak{H}}} \left |(d\psi )_{\langle gh~g\rangle}\right |^2 
=\frac{1}{GH} \langle\hat{\mathcal{P}}^2\rangle_{\psi}$. 
The latter relation highlights the connection between the
combinatorial and the group theoretical definition of momentum.  Indeed $(d\psi)_{\langle g~gh\rangle} = f_{gh}-f_{g}$ is the $g$-th component of the 
function $\displaystyle{\Delta_{h} \psi = \sum_{g} \partial_{\langle g~gh\rangle}\psi}$, 
whose Fourier components are $\left(\Delta_{h}\psi\right)_{\chi(h)}= p_{\chi(h)} \, \psi_{\chi}$: in the case of an infinite group with infinitely  many inequivalent irreducible representations (like most Lie Groups) a one-to-one
mapping is thus established between $E(X)$ and $V(\widehat{X})$ and the two momentum 
operators become simply proportional to each other. 
Using now general, well established inequalities \cite{velasquez}, one has:
\begin{equation}\label{heisy}
\langle\mathcal{Q}^2\rangle \langle\hat{\mathcal{P}}^2 \rangle \geq \frac{1}{4H} \; . 
\end{equation}
Even if (\ref{heisy}) is not the strictest possible inequality which can be obtained, nevertheless it appears quite  
similar to the standard physical Heisenberg uncertainty inequality. This is suggestive, 
because we are not dealing with actual positions and momenta, nor with their corresponding operators as usually introduced in quantum mechanics, but only with their abstract 
analogue over $X$, nothing more than rough notions of position and direction. Nevertheless this is sufficient to obtain the inequality (\ref{heisy}); 
a hint of the deep ultimate origin of the fundamental Heisenberg uncertainty in quantum 
mechanics.


The introduction of a precise notion of time evolution is now required for the consistent 
construction of a dynamical system. Ambient space will be the given graph $X$, but also time 
will be discretized, in order to supply a fully covariant description.  Furthermore, a discrete time is naturally endowed with partial ordering, provided by its integer labelling; in other words, there 
exists an intrinsic set-theoretical arrow of time, associated with the causal ordering 
of events in special relativity, but also with the algebraic time-flows that one can 
construct in generally covariant theories (such as the thermal time of ref. \cite{connes}).

The  state function of the system will  now be defined over a larger graph, 
called the \emph{time expansion} $\tau(X)$, obtained joining different copies of the 
original graph $X$, in accordance with the scheme induced by the dynamics;
each vertex of the time expansion will be denoted by a double label, $(v,t_i)$, in which 
the first entry, $v\in V(X)$, will once more give the position while the other, $t_i$, 
will have the meaning of temporal coordinate. At time $t_i$, the state function will then 
be written as:
\begin{equation}
\psi_i = \sum_{v\in V(X)} \psi_{(v,t_i)}\, a_{(v,t_{i})} \; , 
\end{equation}
Let focus, for example, on random walks and quantum evolutions and let assume a discrete spacetime model, where infinitely many edges of the type $\langle (v,t_i)(v',t_{i+1}) \rangle$, $i\in 
{\mathbb{N}}$, are introduced for each edge $\langle vv'\rangle$ in $E(X)$. The graph 
endomorphisms generated by vector fields of the form: 
\begin{equation}
{\bf E} = \sum_{\langle vv'\rangle\in E(X), i} E_{\langle (v,t_{i+1})(v',t_i)\rangle} 
\partial_{\langle (v,t_{i+1})(v',t_i)\rangle} \; , 
\end{equation}
will generically describe the system evolution, but other constraints must be respected 
if the resulting dynamics has to be a physically allowed one. In particular, the evolved 
state function $\psi_{i+1}$ must be a linear combination only of the $\{a_{(v,t_{i+1})}\}$, 
because evolution is toward the future (in a sense, the time sequence is completely 
identified with the causal sequence itself). Thus the components along the past basis 
$\{a_{(v,t_{i})}\}$ have to vanish. This happens if and only if the evolution
generator $\textbf{E}$ satisfies the \emph{progressivity  condition}:
\begin{equation}\label{progressivity}
\forall v\in V(X)\; ,\; \sum_{v'\in V(X)} E_{\langle (v,t_{i+1})(v',t_i)\rangle } = 1 \; . 
\end{equation}
Condition (\ref{progressivity}) guarantees that the evolved state function is identical with 
its restriction to the next time-sheet, and there is no state self-overlap in time, thus  
implementing the right arrow of time directly in the evolution law:
$\psi_{i+1} = \left[ \openone + {\bf E} \right ](\psi_{i}) = \sum_{v,v'\in V(X)} E_{\langle 
(v,t_{i+1})(v',t_i)\rangle} \psi_{(v',t_i)}\, a_{(v,t_{i+1})} \; $.
Finally, the progressivity condition can be rephrased as quasi-stochasticity of the 
evolution matrix ${\mathbb{E}}_i$ (in a quasi-stochastic matrix entries are in general 
complex and their sum on each row is equal \mbox{to 1}), whose $vv'$-entry is equal to 
$E_{\langle (v',t_{i+1})(v,t_i)\rangle}$. It is worth noticing that this constraint follows 
only from the existence of a time flow with a precise direction; in the random walk case 
${\mathbb{E}}_i$ has to be stochastic to begin with, being the transition matrix of a
Markov Chain.  Quasi-stocasticity must subsist, however, also in the quantum case, besides 
the conservation of probability, associated usually to the unitarity of the dynamical flow.
A strong connection is then evident between quantum and diffusional dynamics, as suggested 
in the stochastic quantization literature \cite{stocquant1}, \cite{stocquant2}.

Introduce next the temporal derivation operation $\displaystyle{\partial_{\langle \tau_{i} \tau_{i+1} \rangle} = \sum_{v\in E(X)}
\partial_{\langle (v,\tau_{i}) (v,\tau_{i+1})\rangle}}$. The time derivative $\displaystyle{\partial_{\langle \tau_{i}) \tau_{i+1} \rangle}\, \psi}$ can be evaluated directly, for the 
time-dependent state function $\displaystyle{\psi = \sum_i \psi_i}$ of a single particle 
moving on $X$. If $\vec{\psi}_k$ is the column vector of the components of $\psi_k$, it 
is straightforward to write:
\begin{equation}\label{schrodingerwick}
\partial_{\langle \tau_{k} \tau_{k+1}\rangle} \psi = \alpha{\mathbb{H}}_k \cdot \vec{\psi}_k \; , \mbox{\,\,\,\,\,for\,\,\,\,\,} 
{\mathbb{E}}_k = \openone + \alpha{\mathbb{H}}_k \; .
\end{equation}
$\alpha{\mathbb{H}}_k$ can be interpreted as the infinitesimal generator of the evolution matrix 
${\mathbb{E}}_k$, in analogy with what was done in (\ref{infinitesimalgen}). The same 
equation (\ref{schrodingerwick}) describes both the evolution of probability distributions 
and of single particle wave functions and in both cases $\openone + \alpha{\mathbb{H}}$ has to 
be quasi-stochastic. For a random walk, it must be also positive, while for quantum 
evolution unitarity should be ensured.  In this simple model quantum evolution can thus be obtained
as an analytical continuation of a Markov process over the complex plane, that is as a generalized 
random walk with complex transition probabilities; the meaning of such odd entities remains 
far from obvious, but some enlightening $'$physical$'$ remarks can be found in Feynman's 
discussion \cite{feynmannegprob}; moreover complex valued Wiener measures  are already 
known and have been rigorously characterized \cite{hochberg}, \cite{jumarie}. 

If one assumes now by analogy that $\mathbb H$ has the dimensions of an energy, like an actual hamiltonian operator, the constant $\alpha$ must have the form $\alpha=\tau_{0}/\hbar$, where a typical time scale (related to the duration of the elementary time step) has been factored out and $\hbar$ has the dimensions of an action as needed for dimensional consistency. Performing the time variable change $\tau \mapsto -it$, analogous to an inverse Wick rotation,  (\ref{schrodingerwick})  can be recast in the form:
\begin{equation}\label{Schroedinger}
i\hbar\, \partial_{\langle t_{k+1} t_k \rangle} \psi = {\mathbb{H}}_k \cdot \vec{\psi}_k \; ,
\mbox{\,\,\,\,\,for\,\,\,\,\,} 
{\mathbb{E}}_k = \openone -\frac{it_0}{\hbar}{\mathbb{H}}_k \; . 
\end{equation}
Dealing with free particles, a natural choice for the matrix $\mathbb H$ might be proportional to the graph laplacian ${\mathbb{L}}={\mathbb{D}}\, {\mathbb{D}}^{T} = {\mathbb{J}} - {\mathbb{A}} \; $ where ${\mathbb{A}}$, ${\mathbb{D}}$ and ${\mathbb{J}}$ are respectively the adjacency and incidence matrix and the diagonal matrix of valencies of the ambient graph $X$ (see for example \cite{royle}). The incidence matrix provides the matrix form for the momentum operator 
1-form $\omega_{\openone}$ and the laplacian is then a sort of square of the momentum operator. $\mathbb{H}=\frac{\hbar^2}{2m}\mathbb L$, is a symmetric hamiltonian; it is then possible to introduce the vector field  $\mathbf{H}=\sum_{\langle vv'\rangle \in E(X)}\mathbb{H}_{v'v}\partial_{\langle(v',i)(v,i)\rangle}$ and rewrite (\ref{Schroedinger}) as:
\begin{equation}\label{poissonmotion}
\partial_{t_{i},t_{i+1}}\psi_{i}=\frac{1}{i\hbar}\{\mathbf{H},\psi_{i}\}
\end{equation}
The constant $(i\hbar)^{-1}$ appears in front of the Poisson brackets only because of Wick rotation and reasons of dimensional consistency, and not in force of  Bohr Correspondence Principle: there is no quantization to perform simply because the mechanics is quantum-like from the beginning.

Unfortunately, $\openone - it_{0}\frac{\hbar}{2m}{\mathbb{L}}$ is quasi-stochastic but not 
unitary, and the evolution preserves the (normalized) sum of the components of $\psi$ but not 
its 2-norm: the state function can then apparently be considered a complex generalized 
probability distribution but not yet a probability amplitude in the usual sense. Nevertheless, 
let us  introduce a larger time scale $t = n\;t_0$, where $n$ is a positive integer; the 
hamiltonian  is time-independent, then the evolution matrix over the full time interval 
$t$ can be written as a Trotter-Suzuki expansion $\mathbb{E}_n=\left(\openone- 
\frac{i\mathbb{H}t}{\hbar n}\right)^n $, converging for $n$ large enough to ${\mathbb{E}} 
= \exp\left(-\frac{i}{\hbar}{\mathbb{H}}t\right)$. $\mathbb E$ is both quasi-stochastic 
and unitary, then {\em on average}  the norm of $\psi$ is  conserved and at the new time 
scale $t$ the generalized random walker behaves exactly like an ordinary quantum particle. 
Deviations from unitarity are very difficult to observe if the original time scale $t_0$ 
is very small: for instance, if $t_0$ is taken of the order of the Planck time (10$^{-43}$ 
$s$), for a negligibly small renormalized time scale $t \sim 10^{-24}$ $s$ (a $'$yoctosecond$'$) 
the relative deviation would be of the order $10^{-19}$.  At a more formal level the same result, 
hinting at the convergence of the dynamical matrix, comes from direct integration of 
(\ref{poissonmotion}):
\begin{equation}
\psi_{t}=\exp\left[-\frac{it}{\hbar}\mbox{Ad}\;\mathbf{H}\right]\psi_{0}
\end{equation}
written in term of the exponential of the adjoint action of the dynamical algebra.
 
Quantum phenomena (oscillation, interference, localization and delocalization) are then 
quite easily achieved either as coarse-grained  behaviours of a generalized diffusion or 
as solutions of a fundamentally classical equation of motion.  A very interesting feature 
is here that tunnelling between sites not directly connected by edges can be observed: 
for instance if the particle is initially localized at one of the endsites of a linear 
chain, after only one renormalized temporal step $t$ there will be an exponentially damped 
but not null probability of finding it at the opposite lead. Of course such tunneling is 
the byproduct of a much faster sublying quasi-stochastic dynamics. The same happens for 
the related quantum probability interference. The example problem studied in this paper 
is little more than a simple toy model and the theory is still far from being rubust enough 
as to describe real world physics; nevertheless sound arguments for the existence of a 
deep-lying connection between space-time discretization, random processes and quantum 
dynamics have been put forward. It may be reasonably hoped that quantizing geometry may 
indeed help to better understand the very meaning of the quantization of physics. Basic 
pillar of such understanding is mimicking non-commutative geometry \cite{CON}, in the context 
of which it was shown that a mathematical framework can be developed where the fundamental object 
of (continuous) geometry is no longer a manifold but an algebra. Since commutative $C$-$*$ 
algebras biuniquely correspond to locally compact topological spaces, all relevant information 
about the topological structure of such spaces is encoded in their algebra of functions. 
This underlying notion is what has allowed us to construct a $'$differential calculus$'$ 
over ambient spaces given by graphs, resorting to the differential structure that can be 
straightforwardly defined over a commutative algebra.

\end{document}